\documentclass[sigconf]{acmart}

\copyrightyear{2026}
\acmYear{2026}
\setcopyright{cc}
\setcctype{by}
\acmConference[L@S '26] {Proceedings of the Thirteenth ACM Conference on Learning @ Scale}{June 29--July 3, 2026}{Seoul, Republic of Korea.}
\acmBooktitle{Proceedings of the Thirteenth ACM Conference on Learning @ Scale (L@S '26), June 29--July 3, 2026, Seoul, Republic of Korea}
\acmISBN{979-8-4007-2293-6/2026/06}
\acmDOI{10.1145/3774398.3811603}

\renewcommand{\&}{\string\&}
\usepackage{multirow}
\usepackage{balance}

\AtBeginDocument{%
  }

\setcitestyle{nosort}

\begin{document}

\title{Developing Models of Procedural Skills using an AI-assisted Text-to-Model Approach}


\author{Rahul K. Dass}
\orcid{0009-0008-8869-5287}
\affiliation{%
    \institution{Georgia Institute of Technology}
    \city{Atlanta}
    \state{GA}
    \country{USA}}
\email{rdass7@gatech.edu}

\author{Shubham Puri}
\orcid{0009-0009-9862-3532}
\affiliation{%
    \institution{Georgia Institute of Technology}
    \city{Atlanta}
    \state{GA}
    \country{USA}}
\email{spuri62@gatech.edu}

\author{Arpit Khandelwal}
\orcid{0009-0005-2740-6048}
\affiliation{%
    \institution{Georgia Institute of Technology}
    \city{Atlanta}
    \state{GA}
    \country{USA}}
\email{akhandelwal83@gatech.edu}

\author{Xiao Jin}
\orcid{0000-0002-5423-0982}
\affiliation{%
    \institution{Georgia Institute of Technology}
    \city{Atlanta}
    \state{GA}
    \country{USA}}
\email{xjin96@gatech.edu}

\author{Ashok K. Goel}
\orcid{0000-0003-4043-0614}
\affiliation{%
    \institution{Georgia Institute of Technology}
    \city{Atlanta}
    \state{GA}
    \country{USA}}
\email{ashok.goel@cc.gatech.edu}



\begin{abstract}
Scalable AI tutoring for procedural skill learning requires structured knowledge representations, yet constructing these representations remains a labor-intensive bottleneck.
This paper introduces a new LLM-assisted text-to-model (TTM) methodology that transforms instructional materials into schema-complete Task-Method-Knowledge (TMK) models through ontology-constrained prompting and template-based generation, automating structural scaffolding while preserving expert oversight. 
Applied to a graduate-level online AI course, the methodology produced 23 TMK models---enabling full-course coverage for Ivy, a deployed AI coach that relies on TMK models to support learners' procedural understanding, for the first time. 
AI-assisted authoring reduced expert modeling time by 50-70\% while producing structurally valid and highly reproducible models.
We evaluate structural validity, semantic alignment, reproducibility, and refinement effort to characterize authoring scalability.
Results indicate that the TTM methodology substantially lowers the cost of constructing structured procedural representations, making course-wide deployment of structured AI tutoring systems practically feasible.
\end{abstract}




\maketitle

\section{Introduction} \label{sec1}
Online learning environments increasingly integrate generative AI-based intelligent tutoring systems (ITS) to provide interactive support for diverse learner populations \cite{rodrigues2025systematic,letourneau2025systematic}.
In video-based courses, learners frequently seek clarification about procedural skills.
For example, how to implement a method or why a step is valid \cite{koedinger2006cognitive,koedinger2007exploring}.
While large language models (LLMs) can generate fluent explanations \cite{patel2023llm-math-explanations,dong2024llm-education-review}, reliable support for procedural skill learning requires explicit representations of task structure, skill mechanisms (e.g., sequence of actions and logical data conditions), and domain-specific concepts and relations --- properties that are difficult to capture through prompt engineering or retrieval alone.
Recent hybrid approaches that combine structured knowledge models with generative synthesis offer a promising path toward improving explanation reliability in AI-supported learning systems \cite{zhou2025knowledge,ruzimboev2025review}.

One such system is \textit{Ivy} \cite{dass2025ivy,dass2025improving}, a deployed AI coach that combines Task-Method-Knowledge (TMK) models \cite{murdock2008meta,stroulia1999evaluating,goel2017gaia}---structured representations that encode the goals, mechanisms, and domain concepts of a procedural skill---with a LLM to answer learners' ``how-to'' and ``why-based'' questions.
By grounding its explanations in TMK models, Ivy goes beyond retrieval over unstructured text, capturing the causal, teleological, and compositional structure that characterizes procedural skill understanding.
However, Ivy's ability to support a learner's procedural query depends directly on the availability of a TMK model for the skill in question.
At approximately 6--7 hours of expert effort per model \cite{dass2025ivy}, constructing representations for a complete course remains a significant practical bottleneck.

More broadly, the scalability of structured AI tutoring systems depends not only on runtime generation of explanations, but on the feasibility of constructing the underlying knowledge representations.
Prior work in ITS has long identified domain modeling and authoring as a primary bottleneck for scalable deployment: building formal representations of skills requires substantial expert effort and limits coverage across courses and curricula \cite{murray2003overview,sottilare2015design}.
In large-scale online courses that span dozens or hundreds of skills, when each model demands hours of expert design and refinement, full-course coverage becomes impractical.
Under these conditions, the scalability challenge lies less in generative capability and more in scalable knowledge authoring infrastructure.

To address this bottleneck, we present a new text-to-model (TTM) methodology that uses LLMs to semi-automate the construction of TMK models from instructional materials through ontology-constrained prompting and template-based generation.
The methodology automates structural completion tasks such as extracting domain concepts and scaffolding procedural mechanisms while preserving expert oversight for validating causal transitions and failure conditions. 
Our central research question is:
\begin{description}
    \item[RQ:] To what extent can a novel LLM-assisted TTM methodology reduce the expert effort required to construct TMK models of procedural skills, enabling scalable deployment of structured AI tutoring systems?
\end{description}

In a case study using instructional materials from a graduate-level online AI course at Georgia Tech, we applied the TTM methodology to construct 23 TMK models---achieving full-course coverage for Ivy's structured AI coaching for the first time.
AI-assisted authoring reduced expert modeling time by 50–70\% while producing structurally valid and reproducible models, indicating that course-wide structured AI tutoring deployment is practically feasible.

This work makes three contributions: (1) a new LLM-assisted TTM methodology for semi-automated construction of TMK models from instructional materials, (2) an evaluation framework assessing structural validity, semantic alignment, reproducibility, and refinement effort of TTM-generated models, and (3) empirical evidence that the TTM methodology substantially reduces expert authoring time, enabling realistic course-wide deployment of structured AI tutoring systems such as Ivy.

\section{Related Work}

Text-to-model methods bridge the gap between human-readable text and machine-interpretable representations. 
Large language models and neural-symbolic integration have recently accelerated progress in this area, enabling end-to-end systems that convert natural language directly into formal knowledge representations.

\subsection{Authoring Bottlenecks in Intelligent Tutoring Systems}

The authoring bottleneck remains the central obstacle to scaling ITS. 
Murray demonstrated that developing model-tracing tutors required approximately 200--300 hours of expert effort per hour of instruction, with the majority of time spent constructing production rule models \cite{Murray2003}. 
Aleven reported 1:50 to 1:100 development ratios for cognitive tutors even with improved authoring tools, noting that domain experts struggled particularly with formalizing procedural knowledge into executable rules \cite{aleven2016example}.
Van Lehn distinguished between ``inner loop'' authoring (creating individual problem steps and hints) and ``outer loop'' authoring (curriculum sequencing), identifying inner loop knowledge engineering as the critical bottleneck since outer loop tasks could be automated \cite{vanlegn2006}.
These findings establish that the primary cost in ITS development lies in formalizing domain knowledge, not in runtime generation. 
This motivates approaches that automate structural drafting while preserving expert oversight for semantic validation.

\subsection{LLM-Assisted Structured Knowledge Construction}
LLMs have demonstrated strong capability for transforming unstructured text into structured representations conforming to predefined schemas. 
Schema-constrained extraction imposes structural constraints that ensure outputs adhere to specific ontologies or database schemas. 
KnowCoder \cite{li-etal-2024-knowcoder} represents schemas as Python classes in a format LLMs can interpret naturally, enabling systematic extraction of structured information while maintaining semantic richness. 
PARSE \cite{shrimal-etal-2025-parse} improves structured extraction reliability by automatically refining JSON schemas and applying reflection-based guardrails, directly addressing the hallucination and schema-compliance failures common in LLM-based extraction pipelines.

While these approaches demonstrate that LLMs can reliably produce schema-conformant structured outputs, they target general knowledge extraction tasks. 
Applying schema-constrained generation to the authoring of pedagogically structured procedural models, where correctness of causal transitions and goal decomposition matters as much as structural validity, remains an open problem.

\subsection{Evaluation of structured model generation}
Evaluating LLM-generated structured outputs presents distinct challenges beyond traditional NLP metrics. 
Zhang demonstrated that even schema-valid knowledge graphs contain substantial ungrounded facts, with hallucination rates varying by relation type and domain complexity \cite{zhang2025siren}.
Automatic metrics frequently misalign with human judgment and struggle to detect structural inconsistencies \cite{zhang2025siren, chang2024survey}, and structural validity alone does not guarantee domain relevance \cite{gangemi2006modelling, alshahrani2021application}. 
Recent work emphasizes combining automated checks with expert-driven validation to ensure practical usability \cite{gao2025llm} - our evaluation framework follows this design, pairing rule-based syntactic checks with LLM-assisted semantic assessment.

\subsection{Modeling Procedural Skills using Task-Method-Knowledge} \label{sec2:tmk}
Task-Method-Knowledge (TMK) is a structured knowledge representation designed to model procedural skills by making explicit the goals of a skill (\textit{why}), the mechanisms used to achieve those goals (\textit{how}), and the domain concepts and relations that those procedures operate over (\textit{what}) \cite{murdock2008meta,stroulia1999evaluating,goel2017gaia}.
TMK originated in knowledge-based AI research as a representation intended to be both human-interpretable and machine-usable for supporting procedural modeling, analysis, and explanation \cite{chandrasekaran1992task,chandrasekaran1986generic}. 

In contrast to more widely used planning representations such as Hierarchical Task Networks (HTNs), which is used to emphasize goal-directed procedures \cite{erol1994htn,nau2003shop2}, TMK additionally encodes expectations and constraints that help connect decomposition structure to conditions, outcomes, and domain semantics. 
Prior work has argued that TMK-style representations can be more expressive than standard HTN formalisms for capturing procedural knowledge that supports prediction and explanation, making TMK a strong fit for AI in Education settings where systems must explain not only what to do next, but why a step is warranted and when a strategy fails \cite{hoang2005hierarchical,lee2006study}.
TMK models represent a procedural skill through three interrelated components:

\textbf{Task}: defines the skill's goals and subgoals and specifies what counts as success.
Tasks typically include typed inputs and outputs, preconditions (``given''), postconditions (``makes''), and references to one or more methods that can accomplish the goal.
This structure makes teleological organization explicit by linking procedural activity to goal satisfaction criteria. 
A minimal Task schema is shown in Table \ref{tab:task-schema}.

\begin{table}[h]
\caption{Task Component Schema}
\label{tab:task-schema}
\centering
\setlength{\tabcolsep}{4pt} 
\begin{tabular}{p{0.34\columnwidth} p{0.59\columnwidth}}
\toprule
Attribute & Description \\
\midrule
\texttt{name} & Name of the goal or subtask \\
\texttt{description} & Natural language description of the goal or subtask \\
\texttt{inputParameters} & Inputs required to invoke the goal \\
\texttt{outputParameters} & Outputs or effects produced upon completion \\
\texttt{given} & Preconditions that must hold before invocation \\
\texttt{makes} & Postconditions expected after goal achievement \\
\texttt{mechanism\allowbreak Reference} & References to Method components that realize the goal \\
\bottomrule
\end{tabular}
\end{table}

\textbf{Method}: defines how Tasks are accomplished. 
Methods may be represented as organizers - finite-state machines (FSMs) whose states invoke subtasks or atomic operations - or as reusable operations that directly transform world state.
Transitions are governed by logical conditions and may include explicit success and failure paths.
This structure supports causal chaining by making explicit how state changes depend on conditions and actions.
A minimal Method schema is shown in Table \ref{tab:method-schema}.

\begin{table}[h]
\caption{Method Component Schema}
\label{tab:method-schema}
\centering
\setlength{\tabcolsep}{4pt} 
\begin{tabular}{p{0.34\columnwidth} p{0.59\columnwidth}}
\toprule
Attribute & Description \\
\midrule
\texttt{name} & Name of the mechanism \\
\texttt{description} & Natural language description of the mechanism \\
\texttt{inputParameters} & Inputs required for execution \\
\texttt{outputParameters} & Outputs produced \\
\texttt{requires} & Preconditions that must hold before execution \\
\texttt{provides} & Postconditions expected after successful execution \\
\texttt{organizer} & Sequence of states and transitions \\
\hline
\end{tabular}
\end{table}

\textbf{Knowledge}: defines the domain ontology over which Tasks and Methods are expressed, including concepts and relations. 
Knowledge provides the vocabulary and constraints needed for type consistency and for expressing conditions used in task definitions and method transitions.
A minimal Knowledge schema is shown in Table \ref{tab:knowledge-schema}.

\begin{table}[h]
\caption{Knowledge Component Schema}
\label{tab:knowledge-schema}
\centering
\setlength{\tabcolsep}{4pt} 
\begin{tabular}{p{0.34\columnwidth} p{0.59\columnwidth}}
\toprule
Attribute & Description \\
\midrule
\texttt{Concept} & Domain entity type (e.g. List, Element) \\
\texttt{superConcept} & Declares parent-child class relationships \\
\texttt{properties} & Typed attributes of the concept \\
\texttt{Instance} & Specific named instances of a concept \\
\texttt{values} & Property-value bindings for an instance \\
\texttt{Relation} & Links between concepts \\
\hline
\end{tabular}
\end{table}

\begin{figure*}[!h]
    \centering
    \includegraphics[width=0.75\textwidth]{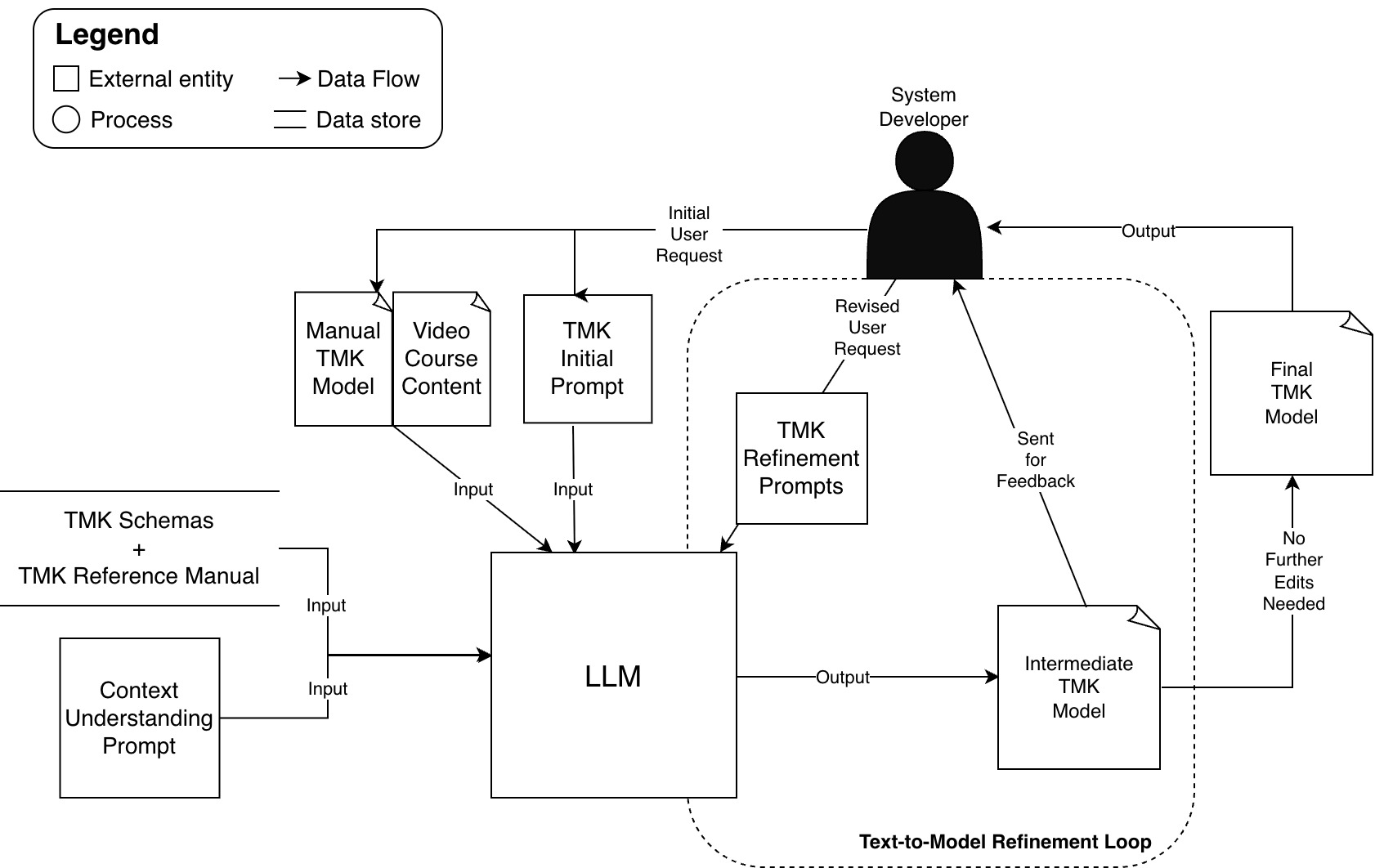}
    \caption{Human-in-the-loop text-to-model architecture illustrating schema-constrained generation and iterative refinement.}
    \label{fig:ttm-arch}
\end{figure*}

For example, a simplified procedural skill such as \textit{Sorting a list} could be represented with a Task that defines the goal (\texttt{produce an ordered list}) and references a sorting method; a Method that captures control structure (e.g., \texttt{iterative insertion}) with transitions guarded by conditions (e.g., \texttt{comparisons}); and Knowledge that defines concepts such as \texttt{List} and \texttt{Element} and relations such as \texttt{greaterThan}.
All three components together encode not only step structure but also the conditions under which steps apply and the criteria for successful completion.

In ``structured AI tutoring systems'', TMK models can serve as structured scaffolds that constrain or guide generative components, grounding explanations in explicit goal structure, state transitions, and domain terminology.
Because TMK models require explicit encoding of decomposition structure, transitions, and domain constraints, constructing them manually can be time-intensive, motivating text-to-model (TTM) methods that aim to semi-automate model drafting while preserving expert oversight.

Taken together, prior work establishes two complementary foundations: LLMs can perform reliable schema-constrained generation, and TMK provides a structured framework for representing procedural skills in AI tutoring.
To the best of our knowledge, no prior work has combined these to address the authoring bottleneck directly; existing text-to-model systems either target general knowledge graphs rather than pedagogically structured procedural models, or operate at runtime rather than at authoring time.
Our TTM methodology fills this gap by applying ontology-constrained LLM generation to produce TMK drafts from instructional materials, embedding expert refinement into the workflow to ensure instructional fidelity at scale.

\section{Text-to-Model Authoring Pipeline}
Constructing structured models of procedural skills has traditionally required manual, expert-driven authoring, involving careful interpretation of instructional materials and explicit encoding of procedural structure, state transitions, and domain constraints.
While expressive, this process is time-intensive and does not scale to full-course coverage.
To address this bottleneck, we introduce a human-in-the-loop text-to-model (TTM) pipeline that leverages large language models to generate structured drafts under explicit representational constraints.
The pipeline operates at authoring time and produces persistent symbolic models rather than runtime explanations.

\subsection{Ontology-Constrained Model Construction}
The TTM pipeline follows an ontology-constrained generation process, illustrated in Figure~\ref{fig:ttm-arch}.
Model developers provide instructional artifacts (e.g., lecture transcripts or course documents) to an LLM preconfigured with formal JSON schemata (\texttt{.schema.json}) specifying required fields, value types, and structural constraints for each TMK component, and reference documentation describing representational semantics\footnote{The full set of TMK schemata and the system prompt used to configure the pipeline are available in the supplementary materials at \url{https://doi.org/10.17605/OSF.IO/NYFPK}.\label{fn:supp}}.

For example, the schemata mandates a \texttt{dataCondition} check on every FSM transition and explicit \texttt{means} references linking Tasks to their realizing Methods, ensuring that structurally critical relationships cannot be omitted from generated drafts.
In this study, we implemented the TTM generation pipeline using Google Gemini~3, configured with ontology-constrained prompts and these JSON schemata, shifting model construction from free-form text synthesis to structured completion under explicit ontological constraints.
The resulting TMK model is represented as a structured JSON artifact conforming to the specified schemata (Section~\ref{sec2:tmk}), enabling static validation, programmatic inspection, and downstream use by tutoring systems.

\subsubsection{TMK Modelling System Prompt Overview}
To ensure that the configured Gemini-3 LLM instance within the TTM pipeline generates specification-compliant Raw TMK drafts, the LLM was initialised with a comprehensive system prompt and a suite of reference materials such as TMK reference documentation and three JSON schemata (one per component).
The system prompt covers five components: (1)~\textit{Objective and Setup}, defining the task of generating schema-faithful TMK JSON files from course materials without invented content; (2)~\textit{Core Philosophy}, enforcing strict fidelity to source material over any pretraining or external knowledge; (3)~\textit{Modeling Instructions}, specifying construction rules for each TMK component including mandatory FSM patterns, transition conditions, and decomposition depth; (4)~\textit{Mandatory Patterns}, enumerating structural requirements present in every generated model (e.g., explicit \texttt{Done} and \texttt{Fail} states, non-trivial \texttt{dataCondition} guards); and (5)~\textit{Step-by-Step Workflow}, directing the LLM to build Knowledge first, then Task, then Method, and validate the final output against the success assertion. 
The full prompt is available in our supplementary materials (see footnote~\ref{fn:supp}).

\subsection{Iterative Expert Refinement}
Generated drafts undergo at least two expert reviews to assess correct causal ordering of procedural steps, completeness of failure conditions, alignment with instructional intent, and semantic consistency with domain concepts.
Identified deficiencies are translated into targeted refinement prompts, and revised drafts are generated iteratively. The process continues until structural and semantic criteria are satisfied.
This workflow establishes a division of labor: the LLM accelerates structural completion tasks such as extracting domain concepts and scaffolding procedural mechanisms, while experts ensure correctness of transitions, constraints, and pedagogical fidelity.

\subsection{Authoring Workflow}
The pipeline is instructor-facing and requires minimal interaction.
Instructional materials are provided, an initial structured draft is generated, and iterative refinement produces a deployment-ready model.
Across 23 procedural skills in a graduate-level online course, this ontology-constrained authoring process reduced expert modeling time by 50–70\% while producing structurally valid and reproducible models, as reported in Section \ref{sec:5.3}.

\subsection{Human-in-the-Loop Refinement: A Running Example}
\label{sec:3.4}

To illustrate the pipeline in practice, we follow a UX persona through the TMK modelling process. 
Organic chemistry professor Aqsa uploads the IUPAC nomenclature guide she uses in her course to the TMK Modelling Gem with the prompt: \textit{``Go through the attached PDF and draft a TMK model of systematic nomenclature of organic compounds.''} 
The system returns three JSON outputs - Task, Method, and Knowledge — which are structurally valid, as confirmed by the static validator, and coherent at a high level. 
However, making them pedagogically reliable requires manual refinement. 
It takes Aqsa approximately 1.5 hours to refine the model to maturity using her domain expertise.
Working with raw JSON syntax and tracking FSM logic textually imposes a significant cognitive load, particularly as she is not a programmer.

Her refinements are concentrated in the Method component, where high-level descriptions lack atomic operations and bypass transitions, and in the Knowledge component, where seniority relations are conflated and some concepts are missing. 
The Task component is comparatively strong, though it misapplies conditions that fail for edge cases like unsubstituted alkanes. 
Across all three components, error propagation is a recurring pattern: a mis-specified Task requires corresponding corrections in the Methods and Knowledge that reference it. 
The full commit-level diff for this example is available in the supplementary materials at \url{https://doi.org/10.17605/OSF.IO/NYFPK}.
Overall, the example illustrates the intended division of labor: the LLM provides strong structural scaffolding that substantially reduces initial effort, while the domain expert's input remains indispensable for ensuring accuracy, completeness, and pedagogical reliability.

\section{Evaluation} \label{sec4:eval}
To assess our research question - whether AI-assisted TTM methods reduce the manual effort required to construct explicit procedural skill representations at scale - we conduct a structured comparison between: (1) \textit{Raw TMKs}: generated exclusively by the LLM (no human intervention), and (2) \textit{Refined TMKs}: models produced after expert iteration and verification.

This distinction is central. 
Raw TMKs reflect the capability of AI-assisted generation. 
Refined TMKs represent deployment-ready artifacts after expert oversight.
By comparing these two sets of TMK models and relating them to prior expert-crafted models, we quantify both structural quality and refinement effort.
Our evaluation is separated into three complementary categories:
\begin{itemize}
    \item Syntactic (rule-based) metrics,
    \item Semantic (similarity-based and LLM-assisted) metrics, 
    \item Expert refinement effort and qualitative analysis. 
\end{itemize}
Together, these dimensions allow us to assess structural validity, instructional fidelity, semantic reasoning quality, stability, and the reduction in manual authoring effort.

\subsection{Syntactic Evaluation: Structural Integrity and Instructional Grounding}

We first evaluate structural properties of TMK JSON models using a static analysis framework aligned with the TMK syntax and semantics.
These metrics assess whether AI-generated drafts are formally valid, internally coherent, and instructionally grounded prior to expert refinement.

\textbf{Instructional Alignment}. A central concern in AI-assisted modeling is whether extracted procedural concepts reflect the vocabulary and structures present in the source instructional materials, rather than generic abstractions derived from model pretraining.
We therefore introduce Instructional Alignment, a lexical grounding metric that measures consistency between Task-Method-Knowledge concepts in the generated TMK model and terminology appearing in the corresponding lesson transcripts.

Using a token similarity ratio that tolerates minor morphological variation, this measure rewards instructor-consistent terminology while penalizing conceptual drift.
Instructional Alignment bridges structural correctness and semantic fidelity by verifying that extracted entities are both formally represented in TMK syntax and grounded in the instructional materials, reducing the risk of instructionally misaligned abstractions.

\textbf{Structural Binding}. We assess the interconnections between each component of a TMK model (Task-Method-Knowledge) across three dimensions: (1) \textit{Task-Method} binding: proportion of tasks with at least one associated method, (2) \textit{Method–Knowledge} binding: percentage of method parameter types defined in the Knowledge graph, and (3) \textit{Task–Knowledge} binding: consistency of input/output types between task definitions and knowledge concepts.

These checks ensure that models are not merely syntactically valid JSON artifacts, but coherent procedural structures with consistent cross-component references.

\textbf{Procedural Heuristics}. We additionally measure structural properties indicative of procedural completeness: (1) \textit{Guard Logic}: percentage of transitions containing non-trivial conditions, (2) \textit{Failure Modeling}: presence of explicitly defined failure states when instructionally relevant, and (3) \textit{Hierarchy Depth}: maximum invocation depth in the task–method decomposition tree.
These heuristics operationalize whether a model meaningfully encodes causal transitions, hierarchical structure, and edge-case reasoning, central properties of procedural skill representation.

For example, Guard Logic captures transitions governed by substantive data conditions rather than trivial existence checks, such as:
\begin{verbatim}
{ 
  "sourceState": "SNS_S5", 
  "targetState": "SNS_Done", 
  "dataCondition": "stereochemistryAssigned(config) 
                   || noStereoPresent(config)" 
}
\end{verbatim}

Similarly, Failure Modeling detects explicitly defined failure states, e.g.:
\begin{verbatim}
{ 
  "name": "SLM_Fail", 
  "goalInvocation": { 
    "goalReference": "FailureGoal", 
    "type": "task", 
    "actualArguments": [] 
  } 
}
\end{verbatim}

Hierarchy Depth reflects multi-level task decomposition, such as:
\begin{verbatim}
Top-level Task: SystematicNamingSolution
  - Method: PrincipalGroupMechanism
    - Subtask: ParentNamingMechanism
      - Atomic Operation: NameHydrideChainOrRing
\end{verbatim}
This illustrates an n-level decomposition, evidencing hierarchical structuring of procedural knowledge.

\subsection{Semantic Evaluation: Similarity-Based Comparisons} \label{sec4:semantic_eval}
Syntactic correctness does not guarantee semantic adequacy. 
We therefore conduct structured pairwise similarity analyses across three comparisons for a given procedural skill:
\begin{itemize}
    \item \textbf{Raw TMK vs. Expert-Crafted Baseline}. We compare Raw TMKs to manual expert-crafted models which serve as a reference approximation of ``ground truth''. 
    This measures the initial quality of AI-generated drafts before refinement.
    \item \textbf{Raw TMK vs. Refined TMK}. We compare Raw TMKs to their corresponding Refined versions. 
    The magnitude of change between these states indicates the degree of structural and semantic adjustment required and serves as an indirect proxy for refinement effort.
    \item \textbf{Same Inputs, Two Generations}. We generate Raw TMKs multiple times with identical inputs and prompts and compute similarity across sessions. 
    This measures generation stability and evaluates whether ontology-constrained prompting reduces variance across runs.
\end{itemize}

For each comparison, we compute cosine similarity at the level of Task, Method, and Knowledge components. 
Because per-field scores are not symmetric, we compute similarity in both input permutations and report averaged scores.
Results are aggregated per-JSON model and per-field.
These comparisons allow us to assess draft quality, refinement magnitude, and generative stability in a principled and transparent manner.

\subsection{LLM-Assisted Semantic Evaluation}
Certain dimensions of procedural reasoning such as causal chaining between states or 
teleological clarity are not fully captured by rule-based structural checks.
Therefore, we used an ``LLM-as-judge'' evaluation approach to compare TMK models against 
their corresponding lesson transcripts.
A GPT-4o judge was presented with each TMK JSON alongside the source transcript and 
asked to assess three dimensions on a 1--5 Likert scale (normalized to $[0,1]$):
\begin{itemize}
    \item \textbf{Causal Chaining}: Whether state transitions reflect domain-specific mechanisms rather than abstract placeholders.
    \item \textbf{Teleological Linkage}: Clarity of goal decomposition and Task-Method alignment.
    \item \textbf{Procedural Fidelity}: Representation of algorithmic structure, including loops and edge cases described in the lesson.
\end{itemize}
Full scoring criteria and prompts are provided in the supplementary appendix (Appendix~B): \url{https://doi.org/10.17605/OSF.IO/NYFPK}.
We use these scores comparatively (Raw TMK vs.\ Refined TMK) rather than as absolute 
measures of correctness.

\subsection{Expert Refinement Time and Effort}
The central bottleneck in a human-in-the-loop pipeline is refinement time.
For each procedural skill, expert developers recorded the time required to transform a Raw TMK model into a deployment-ready\footnote{Deployable TMK models are used by an AI tutor to provide structured procedural explanations with actual learners.} Refined TMK model.
These measurements were compared against prior estimates of fully manual TMK construction time, obtained via self-reported developer logs from an earlier phase of the project prior to any AI-assisted tooling.
These estimates align with a previously reported baseline of approximately seven hours per skill~\cite{dass2025ivy}, which encompassed reviewing instructional materials, constructing the model from scratch, soliciting at least two peer reviews, and incorporating feedback.
Modeling time reduction is computed as:
\begin{equation}
    \text{Reduction} = \frac{\text{Manual Time} - \text{Refinement Time}}{\text{Manual Time}}
\end{equation}
This directly addresses the scalability criteria in our RQ (Section~\ref{sec1}): whether AI-assisted TTM meaningfully lowers the human cost of constructing structured procedural representations.

\subsection{Refinement Areas.}
Finally, we qualitatively analyze recurring refinement patterns across skills.
These include missing or incomplete guard conditions, under-specified failure states, and teleological misalignment between Tasks and Methods.
Such patterns reveal both the strengths and limitations of our TTM methodology.

\section{Results}

\subsection{Structural Validity and Instructional Grounding}
We developed 24 TMK models using the TTM methodology and evaluated them using the framework described above.
Twenty-three models represent skills from an online AI course that serves our primary case study, where structured AI tutoring will be deployed; one additional model from Chemistry was included to probe preliminary cross-domain applicability.
We treat this cross-domain case as exploratory rather than evidence of domain-agnostic generality.

All generated TMK artifacts achieved syntactic validity.
Every model conformed to the TMK JSON schemata with no parsing failures, missing fields, or type violations.
This indicates that schema-complete template scaffolding reliably produces structurally valid models, minimizing low-level correction during refinement.

\begin{table}[h]
\caption{Comparative evaluation of Raw and Refined TMK models. Semantic scores are normalized to $[0,1]$.}
\centering
\small
\begin{tabular}{l l c c c}
\hline
Category & Metric & Raw & Refined & Abs.\ Diff \\
\hline
\multirow{1}{*}{Instructional} 
 & Alignment Score & 0.67 & 0.82 & +0.15 \\
\hline
\multirow{3}{*}{Structural} 
 & T-M Binding & 0.45 & 0.95 & +0.50 \\
 & M-K Binding & 0.14 & 0.35 & +0.21 \\
 & T-K Binding & 0.28 & 0.40 & +0.12 \\
\hline
\multirow{3}{*}{Procedural} 
 & Guard Logic & 0.97 & 0.99 & +0.02 \\
 & Failure Modeling & 0.68 & 0.66 & -0.02 \\
 & Hierarchy Depth & 2.00 & 2.41 & +0.41 \\
\hline
\multirow{3}{*}{Semantic} 
 & Causal Chaining & 0.69 & 0.81 & +0.12 \\
 & Teleological Reasoning & 0.72 & 0.82 & +0.10 \\
 & Procedural Fidelity & 0.67 & 0.79 & +0.12 \\
\hline
\end{tabular}
\label{tab:tmk_eval}
\end{table}

Instructional Alignment improved from 0.67 (Raw) to 0.82 (Refined), indicating that expert iteration strengthens lexical grounding in instructor-specific terminology and reduces conceptual drift (Table \ref{tab:tmk_eval}).
Structural binding improved substantially after refinement, particularly in Task–Method (+0.50) and Method–Knowledge (+0.21) connectivity, reflecting tighter TMK model cohesion between its components after refinement.
Although Raw models scored highly on Guard Logic (0.97), semantic measures show consistent gains in Causal Chaining (+0.12) and Procedural Fidelity (+0.12), suggesting that expert intervention enhances domain-specific specificity rather than merely increasing structural complexity.
Failure Modeling decreased slightly, as generic failure states were removed when not instructionally warranted.

\subsection{Semantic Similarity: Pairwise Comparisons}
Focusing solely on the 23 AI-domain skills from our case study, we report similarity-based results for the three comparisons defined in Section \ref{sec4:semantic_eval}.

\begin{table}[h!]
\caption{Semantic similarity: Expert baselines vs. Raw TMKs, averaged over ($n = 23$) models.}
\centering
\small
\begin{tabular}{l l c c}
\hline
TMK component & Metric & Mean & SD \\
\hline
\multirow{3}{*}{Task}
  & Overall        & 0.79 & 0.08 \\
  & Per-Field      & 0.61 & 0.38 \\
  & Dict-Symmetric & 0.61 & 0.03 \\
\hline
\multirow{3}{*}{Method}
  & Overall        & 0.85 & 0.10 \\
  & Per-Field      & 0.59 & 0.38 \\
  & Dict-Symmetric & 0.61 & 0.06 \\
\hline
\multirow{3}{*}{Knowledge}
  & Overall        & 0.71 & 0.11 \\
  & Per-Field      & 0.29 & 0.32 \\
  & Dict-Symmetric & 0.31 & 0.08 \\
\hline
\end{tabular}
\label{tab:sem_fallvspring}
\end{table}

\textbf{Raw TMK vs. Expert Baselines}.
Across 23 skills, Raw TMKs exhibit strong overall mean cosine similarity with expert-crafted models: Task (0.79), Method (0.85), and Knowledge (0.71), as shown in Table \ref{tab:sem_fallvspring}.
While overall alignment is high, particularly for the Method component, per-field and symmetric scores are substantially lower, especially for Knowledge, indicating variability in fine-grained structural and lexical alignment.
These results suggest that the TTM pipeline meaningfully captures the high-level procedural structure of expert models, while refinement focuses primarily on increasing lexical grounding and relational specificity rather than reconstructing procedural structure from scratch.

\textbf{Generation Stability Under Fixed Inputs}.
To assess the reproducibility of the LLM-based TTM pipeline, we regenerated ($n=9$) models with identical inputs and prompts.
Results in Table \ref{tab:sem_gemiterations} show near-perfect stability across Task, Method, and Knowledge components, with overall similarity scores of 1.00 and per-field scores consistently above 0.90.
This indicates that ontology-constrained prompting and schema templates yield highly reproducible outputs under fixed conditions.
Stability is an important property for scalable authoring workflows, ensuring consistent drafts for instructors and developers.

\begin{table}[h]
\caption{Semantic similarity: Same inputs in two sessions. Averaged over ($n = 9$) models.}
\centering
\small
\begin{tabular}{l l c c}
\hline
TMK Component & Metric & Mean & SD \\
\hline
\multirow{3}{*}{Task}
  & Overall        & 1.00 & 0.00 \\
  & Per-Field      & 0.90 & 0.10 \\
  & Dict-Symmetric & 0.90 & 0.00 \\
\hline
\multirow{3}{*}{Method}
  & Overall        & 1.00 & 0.00 \\
  & Per-Field      & 0.99 & 0.00 \\
  & Dict-Symmetric & 0.99 & 0.00 \\
\hline
\multirow{3}{*}{Knowledge}
  & Overall        & 1.00 & 0.00 \\
  & Per-Field      & 1.00 & 0.00 \\
  & Dict-Symmetric & 1.00 & 0.00 \\
\hline
\end{tabular}
\label{tab:sem_gemiterations}
\end{table}

\subsection{Raw vs. Refined TMKs: Refinement Magnitude and Authoring Effort} \label{sec:5.3}
For a subset of five skills that underwent full expert refinement ($n=5$), we compared Raw TMK drafts to their refined, deployment-ready counterparts.
Overall cosine similarity remained high across components: Task (0.90), Method (0.91), and Knowledge (0.89), as shown in Table \ref{tab:sem_rawrefined}.

\begin{table}[h]
\caption{Semantic similarity: Raw TMKs vs. Refined TMKs, averaged over ($n = 5$) models.}
\centering
\small
\begin{tabular}{l l c c}
\hline
TMK Component & Metric & Mean & SD \\
\hline
\multirow{3}{*}{Task}
  & Overall        & 0.90 & 0.13 \\
  & Per-Field      & 0.78 & 0.28 \\
  & Dict-Symmetric & 0.78 & 0.13 \\
\hline
\multirow{3}{*}{Method}
  & Overall        & 0.91 & 0.06 \\
  & Per-Field      & 0.76 & 0.26 \\
  & Dict-Symmetric & 0.76 & 0.07 \\
\hline
\multirow{3}{*}{Knowledge}
  & Overall        & 0.89 & 0.05 \\
  & Per-Field      & 0.60 & 0.37 \\
  & Dict-Symmetric & 0.66 & 0.28 \\
\hline
\end{tabular}
\label{tab:sem_rawrefined}
\end{table}

These results indicate that Raw TMK drafts generally capture the intended high-level conceptual structure of each skill.
However, lower per-field and semantic similarity scores, particularly for the Knowledge component, reveal variability at finer structural levels.
This pattern suggests that refinement consists primarily of localized but structurally significant edits rather than full model reconstruction.
Common revisions included tightening guard conditions, decomposing underspecified subtasks, adding missing transitions, and correcting type bindings to better align with instructional intent.

To quantify reduced authoring effort, the same five models were timed during expert refinement.
Prior to TTM adoption, manual TMK construction averaged 6–7 hours per skill.
Under the TTM workflow, refinement required a mean of 1.9 hours (median 2.0 hours; SD 0.42 hours), representing roughly a two-thirds reduction in expert modeling time.

Together, the high structural similarity between Raw and Refined models and the substantial reduction in refinement time indicate that TTM meaningfully lowers the human effort required to construct structured procedural representations.
Expert labor shifts from drafting core structure to targeted semantic correction, supporting scalability while preserving instructional fidelity.

\subsection{Refinement Patterns and Cross-Domain Illustration}
Across Refined TMKs, we observe consistent refinement patterns that clarify where expert effort is concentrated:
\begin{itemize}
    \item Strong structural scaffolding: Raw drafts reliably satisfy TMK structure and provide usable high-level decompositions.
    \item Incomplete procedural decomposition: Multi-step operations may be collapsed or underspecified, requiring deeper hierarchical breakdown.
    \item Guard conditions and edge cases: Preconditions are often present but require tightening to correctly handle bypass cases and boundary conditions.
    \item Error propagation across components: Incomplete task decompositions frequently propagate to Method finite-state machines (FSMs) and Knowledge assertions.
    \item Retrieval artifacts: Section and figure references from source materials may appear and must be removed.
    \item Usability considerations: Editing JSON and maintaining FSM consistency may introduce cognitive load, particularly for non-programmers.
\end{itemize}

To probe preliminary cross-domain applicability, we applied the pipeline to an organic chemistry procedural skill (IUPAC nomenclature).
Consistent with AI-domain results, the Raw draft was structurally valid and provided a strong initial decomposition.
Refinement focused on correcting domain-specific edge cases, tightening preconditions, and adding missing hierarchical and relational constraints required by the procedure.

We treat this single non-CS example as an exploratory cross-domain probe rather than evidence of full domain-agnostic modeling. Broader cross-domain validation remains future work.

\section{Discussion}

Our findings demonstrate that LLM-assisted generations substantially accelerate the construction of TMK ontological models.
Across 24 models, the TTM pipeline reduced authoring time from approximately 6--7 hours per skill~\cite{dass2025ivy} to an average of about two hours, representing roughly a two-thirds reduction in expert effort.
This reduction is not merely quantitative; it reshapes the authoring process by allowing experts to focus on higher-order semantic refinements rather than structural scaffolding.

Evaluation results show that Raw TMK drafts consistently capture high-level procedural structure and content.
The Task and Method components exhibit strong overall semantic similarity to expert-crafted models, while syntactic validity is achieved in all cases through schema-constrained prompting.
Structural binding and instructional alignment improve substantially after refinement, indicating that expert effort concentrates on tightening relational consistency and grounding terminology in lesson-specific language rather than reconstructing models from scratch.

Interestingly, the TTM methodology consistently produced components in the order Knowledge $\rightarrow$ Task $\rightarrow$ Method.
Although not explicitly required, this ordering reflects the epistemic structure embedded in the TMK specification: domain knowledge informs cognitive tasks, which in turn decompose into executable methods. 
The generation sequence suggests that ontology-constrained prompting implicitly mirrors the hierarchical logic of procedural skill representation.

Despite these strengths, expert refinement remains essential.
The most significant editing effort is devoted to decomposing tasks into atomic operations, tightening guard logic, and correcting FSM transitions.
Occasional omissions in edge-case handling and relational constraints highlight current limitations of LLM abstraction.
However, because syntactic correctness is guaranteed by schema-complete templates, refinement effort shifts from structural drafting to semantic precision.

The depth of hierarchical decomposition (i.e., Task $\rightarrow$ Method $\rightarrow$ ... $\rightarrow$ Task $\rightarrow$ Knowledge) and whether multiple solution methods can be applied to solve a single Task are influenced by the expert's representational choices, not solely by objective properties of the underlying domain.
The TTM pipeline currently inherits abstraction biases from both instructional materials and LLM priors. 
Future work must investigate mechanisms for controlling representational granularity and supporting alternative method encodings.

This division of labor aligns with a System 1 / System 2 framing \cite{kahneman2011thinking}: rapid structural completion by the LLM followed by deliberate expert validation.
More broadly, the TMK schemata constrain LLM output during authoring in much the same way that TMK models constrain LLM inferencing during tutoring \cite{murdock2008meta,dass2025ivy,lum2025designing}.
In both cases, unstructured information is transformed into explicit relational structure, preserving domain relationships while making them computationally operable.

Taken together, these findings directly address our research question from Section \ref{sec1}.
AI-assisted TTM does not eliminate expert involvement, but it meaningfully reduces the manual effort required to construct structured procedural representations suitable for scalable AI tutoring systems.

\section{Limitations}

Despite significant gains in authoring efficiency, several limitations remain.

First, although hallucinations were relatively rare, occasional conceptual errors propagated across components.
For example, a mis-specified Task often required corresponding corrections in the Method and Knowledge components.
Such cascading errors highlight the fragility of automated ontology construction and reinforce the necessity of expert validation prior to deployment.

Second, decomposition depth was consistently limited in Raw TMK drafts. 
Although Tasks were typically broken into subtasks, deeper hierarchical decomposition---to subsubtasks and atomic operations---often required expert intervention.
This constrains the granularity of procedural fidelity achievable through automated drafting alone.

Finally, the quality and genre of instructional materials likely influence the quality of Raw TMKs. 
Because our case study relied primarily on lecture transcripts, it remains unclear how the pipeline would perform when applied to textbooks, standards documents, or peer-reviewed articles, which may differ in structure, completeness, and procedural explicitness.

\section{Future Work}
Future work proceeds along three complementary directions: scalability, generality, and usability.

First, while this study demonstrates significant reductions in authoring effort within a single course domain, broader curricular coverage remains a central goal.
Extending the TTM pipeline to additional courses/domains and instructional contexts will allow us to evaluate how well ontology-constrained authoring scales across varied procedural skill types.
In parallel, system-level evaluations of structured AI tutoring will assess how improvements in model authoring translate into explanation quality and learner outcomes, providing an end-to-end view of impact.

Second, generality remains an open research question. 
While preliminary cross-domain modeling, such as the IUPAC nomenclature example in organic chemistry, suggests transferability, systematic validation across domains is needed. 
Procedural skills vary substantially in abstraction, evaluation criteria, and representational structure; skills involving interpretive judgment, creative design, or loosely structured problem-solving may challenge ontology-constrained representations and require extensions to the framework.

Finally, usability improvements are essential for broader adoption. 
The current workflow requires direct JSON editing and FSM inspection, which can impose significant cognitive load on non-programmers, as illustrated in the running example in Section \ref{sec:3.4}.
Developing graphical or schema-aware editing interfaces would preserve expert control while lowering the barrier to entry, making scalable structured authoring feasible for instructors without formal AI training.

\section{Conclusion}
This paper introduced a new LLM-assisted text-to-model (TTM) methodology for semi-automated construction of Task-Method-Knowledge (TMK) models from instructional materials.
By combining ontology-constrained prompting with schema-complete templates and embedding iterative expert review into the workflow, the methodology achieves substantial reductions in authoring time while preserving representational fidelity. 
High-level procedural abstractions---tasks, methods, logical constraints, domain knowledge structures---were reliably captured, with expert oversight concentrated on hierarchical composition depth, causal chaining, and FSM logic rather than structural reconstruction.

The dual-process framing of the methodology, rapid structural completion (System 1 thinking) followed by deliberate expert validation (System 2), proved effective in balancing efficiency with accuracy.
Schema scaffolding eliminated malformed outputs, semantic similarity analyses revealed strong alignment between Raw TMK drafts and expert models, and LLM generation stability ensured reproducibility under fixed inputs.
Together, these properties support scalable authoring workflows suitable for structured AI tutoring systems in large online courses.

The practical significance of this contribution is most directly realised through Ivy \cite{dass2025ivy,dass2025improving}, a deployed AI coach that relies on TMK models to generate structured procedural explanations for learners in a graduate-level online AI course. 
The TTM methodology enabled full-course TMK model coverage for Ivy for the first time, supporting 115 learners in Spring 2026, making course-wide structured AI coaching practically feasible.
Ongoing work is assessing the extent to which Ivy's TMK-grounded explanations support measurable learning outcomes, providing an end-to-end view from knowledge authoring to student learning.

At the same time, important limitations remain.
Occasional hallucinations, error propagation across components, and depth-limited decomposition underscore the continued necessity of expert oversight.
Broader cross-domain evaluation is required before stronger claims of generality can be made, and improved authoring interfaces will be essential for adoption by non-programmers.
AI assistance does not replace expert modeling; rather, it shifts expert effort toward higher-value semantic refinement.

Looking forward, the TTM methodology offers a foundation for extending structured TMK model coverage across courses and institutions, exploring alternative levels of knowledge abstraction, and developing accessible tooling to democratize ontology-based authoring.
Situating generative AI within an explicit, expert-guided, ontologically constrained workflow advances a practical path toward scalable, reliable neuro-symbolic AI tutoring systems capable of supporting procedural skill learning at scale.

\begin{acks}
This research has been supported by US NSF Grants \#2115232 and \#2247790 to the National AI Institute for Adult Learning and Online Education (aialoe.org).
We thank Dr. Spencer Rugaber of Georgia Tech's Design Intelligence Laboratory for his invaluable guidance on TMK model development.
We also thank Thomas Bowlin, Zebing Li, and John K Hall for constructing TMK models of procedural AI skills taught in a graduate-level Knowledge-based AI course at Georgia Tech.
Lastly, we thank Umm Hafsa for her subject matter expertise regarding the IUPAC Blue model and Hiba Ambreen for her helpful non-domain-expert feedback identifying expert blindspots in the UI evaluation.
\end{acks}


\bibliographystyle{ACM-Reference-Format}
\balance
\bibliography{references}


\end{document}